\documentclass[twocolumn]{revtex4}
\usepackage{graphicx,epsfig,float}
\usepackage{amsmath}
\usepackage{amsfonts}
\usepackage{physics}
\usepackage{fancyhdr}
\usepackage{bm}
\begin{document}

%%%%%%%%%%%%%%%%%%%%%%%%%%%%%%%%%%%%%%
%%%%%%%%%%%%%%%%%%%%%%%%%%%%%%%%%%%%%%

%\pagestyle{fancy}
%\lhead{\bf Manipulating Quantum Spins by a Spin-Polarized Current: An Approach Based Upon $\mathcal{PT}$-Symmetric Quantum %Mechanics}
%\rhead{Aleix Bou Comas, Eugene Chudnovsky and Javier Tejada}
%%%%%%%%%%%%%%%%%%%%%%%%%%%%%%%%%%%%%%

\title{ Manipulating Quantum Spins by a Spin-Polarized Current: An Approach Based Upon $\mathcal{PT}$-Symmetric Quantum Mechanics}
\author{Aleix Bou Comas\textsuperscript{1}, Eugene M. Chudnovsky\textsuperscript{2} and Javier Tejada\textsuperscript{1}}
\affiliation{\textsuperscript{1}Facultat de F\'isica, Universitat de Barcelona, Diagonal 645, 08028 Barcelona, Spain.\\
\textsuperscript{2}Physics Department, Lehman College and Graduate School, The City University of New York,
	250 Bedford Park Boulevard West, Bronx, New York 10468-1589, USA.}

\begin{abstract}
We propose a quantum processor based upon single-molecule magnets and spin transfer torque described by $\mathcal{PT}$-symmetric quantum mechanics. In recent years $\mathcal{PT}$-symmetric Hamiltonians have been used to obtain stability thresholds of various systems out of equilibrium. One such problem is the magnetization reversal due to the spin transfer torque generated by a spin-polarized current. So far the studies of this problem have mostly focused on a classical limit of a large spin. In this work we are discussing spin tunneling and quantum dynamics of a small spin induced by a spin polarized current within a $\mathcal{PT}$-symmetric theory. This description can be used for manipulating spin qubits by electric currents. 
\end{abstract}

\maketitle

\section{Introduction}

Electronic transport through single-molecule magnets (SMM) has been intensively studied in the past. It allowed probing of spin quantum states of an individual SMM as well as of its nuclear spin states \cite{Wernsdorfer2012,Wernsdorfer2016}. A natural question in the context of quantum computation is whether a spin-polarized current through an SMM could allow manipulation of its quantum spin state. Quantum tunneling of a localized spin between equivalent $|\uparrow \rangle$ and $|\downarrow \rangle$ orientations along the magnetic anisotropy axis has been one of the most consequential recent discoveries in spin physics \cite{Friedman,Tejada,Barbara,MQTbook,Friedman-Review,MM}. It has been widely believed that the observed quantum superpositions of spin states can be utilized in qubits. 

\begin{figure}[h]
	\centering
	\includegraphics[width=\columnwidth]{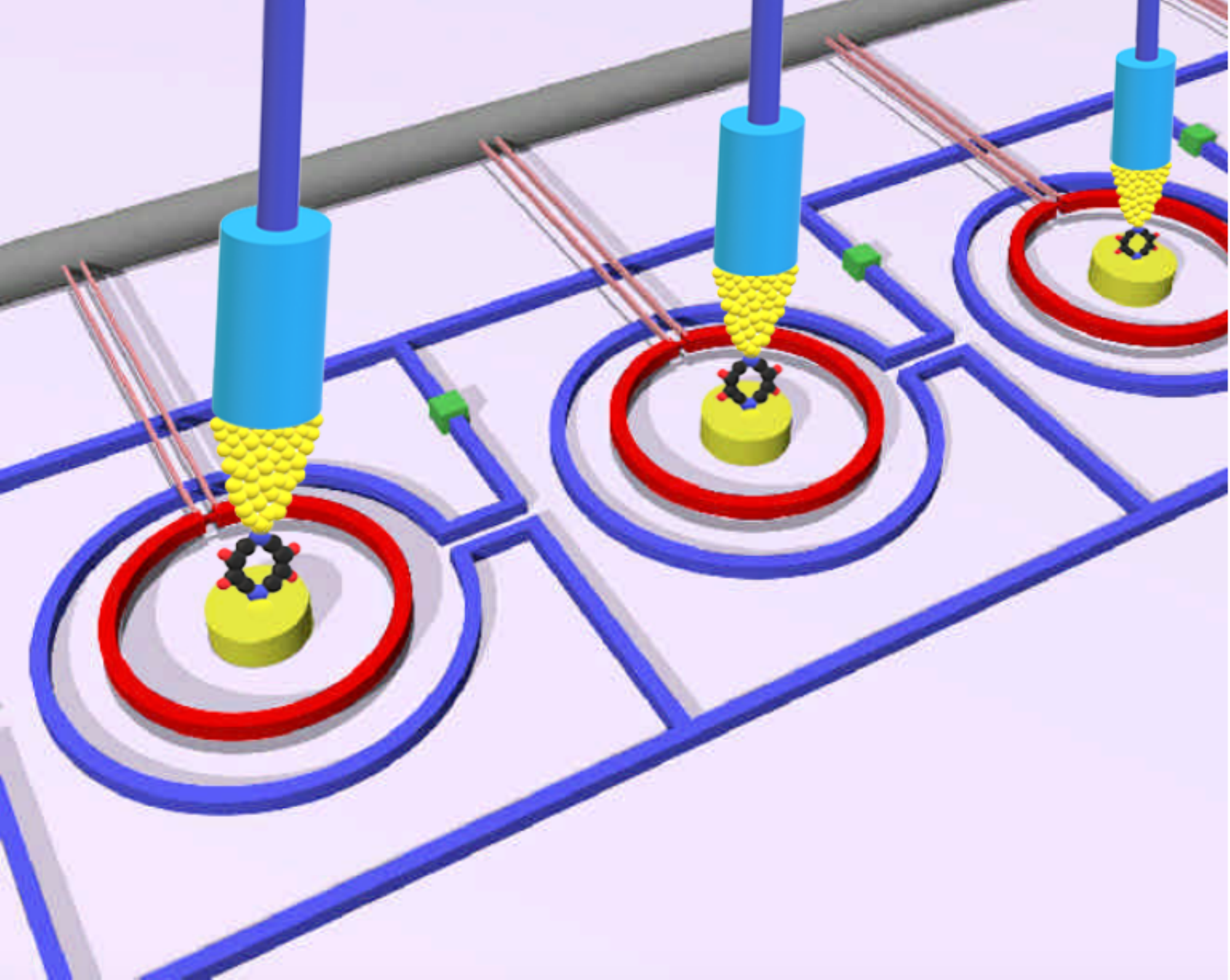}
	\caption{Schematic example of a quantum processor. SMM-based qubits are coupled through nano-SQUID (red) circuits containing Josephson switches (green) as proposed in Ref.\ \cite{Tejada-qubit}. Here we refine this idea by suggesting that quantum states of the SMMs can be manipulated by spin-polarized currents from  ferromagnetic nanopillars (blue), with algorithms based upon $\mathcal{PT}$-symmetric quantum mechanics.}
	\label{fig: device}
\end{figure}
A quantum-gate device based upon SMMs has been proposed by a subset of the authors in Ref.\ \cite{Tejada-qubit}. Magnetic qubits would be arranged in a 1D or 2D lattice and coupled to the superconducting loops of nano-SQUIDs (Superconducting Quantum Interference Device). DiVincenzo criteria: having identifiable qubits with low decoherence, realization of quantum gates, scalability, possibility of the reliable measurement, and workable preparation of quantum states have been discussed. The argument was made that all criteria were within experimental reach but no specific suggestion was made at the time regarding the preparation of quantum states of the qubits. It was noticed that, in principle, it could be done with the help of the external magnetic field but the latter would act on the array of qubits indiscriminately as it would be difficult to localize the external field at the nanoscale. 

Recently a new technology has emerged: single-molecule transistors \cite{Kai,Perrin}, that could provide solution to the problem of selective manipulation of nanoscale magnetic qubits. The schematic structure of the device, incorporating such setups based upon SMMs and metallic ferromagnetic nanopillars as a source of spin-polarized current, is shown in Fig.\ \ref{fig: device}. In this paper we will focus solely on the question that was left out in previous discussions of spin qubits: The possibility of selective manipulation of quantum spin states of SMMs by spin-polarized currents. It will be studied within $\mathcal{PT}$-symmetric quantum mechanics. The specifics of the coupling and measuring magnetic qubits have been discussed at length in Ref.\ \cite{Tejada-qubit} and will not be addressed here. 

One of the most studied non-equilibrium effects in magnetism in recent years has been the magnetization reversal by a spin transfer torque (STT) carried by a spin-polarized current. Initially proposed by Slonczewski \cite{Slonczewski96} and Berger \cite{Berger96} it triggered a wide-spread research on magnetic devices operated by spin-polarized currents \cite{Ralph} that lead to the commercialization of spin-transfer torque random access memory (STT-RAM) devices \cite{Brataas}. An interesting new twist in this area is a recent demonstration by Galda and Vinokur that STT can be described within $\mathcal{PT}$-symmetric quantum mechanics by studying the spectrum of a non-Hermitian $\mathcal{PT}$-symmetric spin Hamiltonian \cite{galda}. In this paper we extend their approach by considering quantum tunneling and time evolution of quantum states of a localized spin, e.g., of an SMM, in the presence a spin-polarized current. 

The $\mathcal{PT}$-symmetric quantum mechanics came into play in the last two decades after it was realized that the Hermitian property of the Hamiltonian mandated by quantum mechanics in order to describe observations can be replaced by a weaker condition of a Hamiltonian having  a $\mathcal{PT}$ symmetry. Since the publication of the seminal papers of Bender, Boettcher, and Meisinger \cite{BB,BBM} the  $\mathcal{PT}$-symmetric theory has been successfully applied to describe non-equilibrium dynamics in non-linear optics and acoustics, Bose-Einsten condensates, superconductors, electronic circuits, etc., see, e.g.,  Ref.\ \cite{Konotop-RMP} for review. The general idea is that a non-Hermitian $\mathcal{PT}$-symmetric addition to the Hamiltonian allows formal generalization of quantum mechanics developed for closed systems to the open systems with a kinetic flow. When the corresponding non-Hermitian term in the Hamiltonian is small it describes the state that is close to equilibrium. This is manifested by a weak perturbation of the eigenstates of the Hamiltonian that leaves the eigenvalues real. The emergence of complex eigenvalues marks the instability threshold that leads to the onset of the dissipative state far from equilibrium. While full conceptual understanding of the foundations of the $\mathcal{PT}$-symmetric quantum theory is far from being settled, its practical value for describing non-equilibrium dynamics of quantum systems is beyond doubt. 

For an integer spin the tunnel splitting $\Delta_m$ between $|m\rangle$ and $|-m\rangle$ states (with $m$ being the magnetic quantum number) is provided by a weak Hermitian perturbation in the Hamiltonian that does not commute with $S_z$. In the absence of other interactions the spin prepared in a state $|m\rangle$ oscillates between $m$ and $-m$ at a frequency $\Delta_m/\hbar$. An interesting question is what happens when the perturbation is not Hermitian but is $\mathcal{PT}$-symmetric, describing interaction of the localized spin (e.g. a spin of a an SMM) with a spin-polarized current. It turns out that on increasing the current the splitting of the spin states computed by the diagonalization of the  $\mathcal{PT}$-symmetric Hamiltonian switches from real to imaginary beginning with the highest-energy states and progressing towards the lowest-energy states. This means that at some critical value of the current the population of one of the states originating from the $|\pm m \rangle$ states begins to grow while population of the other state begins to decrease, effectively taking the spin over the anisotropy energy barrier. The corresponding transition rate rapidly increases on increasing the spin-polarized current and/or temperature.

The paper is organized as follows. STT in a $\mathcal{PT}$-symmetric quantum mechanics is discussed in Section \ref{SST}. Section \ref{coherent} introduces the coherent spin states relevant to the problem. Equations of motion for the expectation value of the spin are derived in Section \ref{motion} to confirm the correspondence \cite{galda} between the spin-polarized current in the phenomenological Landau-Lifshitz-Slonczewski (LLS) equation and the imaginary magnetic field in the non-Hermitian $\mathcal{PT}$-symmetric Hamiltonian. The effect of the spin polarized current on spin tunneling is considered in Section \ref{tunneling}. Section \ref{temporal} is devoted to the temporal evolution of spin states above the stability threshold. Temperature-dependent spin-reversal rate is introduced and studied in Section \ref{reversal}. Section \ref{discussion} contains some estimates and final remarks.

%%%%%%%%%%%%%%%%%%%%%%%%%%%%%%%%%%%%%%%%%%%%%%%%%%%%%%%%%%%%%%%%%%%%%%%%%%%
\section{Spin Transfer Torque in a $\mathcal{PT}$-Symmetric Quantum Mechanics}\label{SST}
\subsection{Spin Coherent States}\label{coherent}
Spin coherent states suitable for the study of non-Hermitian Hamiltonians are \cite{scs}
\begin{equation}
|z\rangle = e^{zS_+}|S, m=-S\rangle.
\end{equation} 
They are holomorphic on the parameter $z$ but not normalized. Applying stereographic projection of the $S^2$ sphere one can use parametrization $z=({1-s_z})^{-1}({s_x+is_y})$, with $s_i={S_i}/{S}$, or $z=e^{-i\phi}\cot({\theta}/{2})$ in terms of the spherical coordinates of points on the sphere of radius 1. Parameter $\bar{z}$ (see below) is defined as $\bar{z}=({1-s_z})^{-1}({s_x-is_y})$ or $\bar{z}=e^{i\phi}\cot({\theta}/{2})$. The state with spin down corresponds to $z=0$, while the state with spin up is represented by $z=\infty$. 

In order to give a physical meaning to these states they must be normalized. Writing
\begin{equation}
1=N^2\langle z'|z\rangle=N^2\langle S, -S|e^{\bar{z'}S_-}e^{zS_+}|S, -S\rangle
\end{equation}
with $N$ being the normalization constant and $z',z$ being the stereographic projections of arbitrarily chosen $|z\rangle$ and $|z'\rangle$, one obtains \cite{scs}
\begin{equation}
1=N^2\sum_{i,j=0}^\infty \frac{\bar{z'}^iz^j}{i!j!} \langle S, -S|S_-^iS_+^j|S, -S\rangle=N^2(1-\bar{z'}z)^{2S}.
\end{equation}
This gives $N=(1-z\bar{z'})^{-S}$ for the normalization constant. 

The expectation value of any operator,  computed with the help of the above coherent states, must be multiplied by $N^2$. For example, the expectation value of the operator $S_z$ is given by
\begin{equation}
\frac{\langle z|S_z|z\rangle}{\langle z|z \rangle}=\frac{\langle S,-S|e^{\bar{z}S_-}S_ze^{zS_+}|S, -S\rangle}{(1+\bar{z}z)^{2j}}  =  S\frac{\bar{z}z-1}{1+\bar{z}z}.
\end{equation}

%%%%%%%%%%%%%%%%%%%%%%%%%%%%%%%%%%%%%%%%%%%%%%%%%%%%%%%%%%%%%%%%%%%%%%%%%%%
\subsection{Equations of Motion for the Spin}\label{motion}

In terms of the coherent spin states the action for the time interval $[0,t_F]$ is given by \cite{scs}
\begin{equation}\label{action}
\begin{array}{ll}
I(z,\bar{z})=&S(\ln(1+\bar{\xi_F}z(t_F))+\ln(1+\bar{z}(0)\xi_I))\\&+\int^{t_f}_0\left(S\frac{\dot{\bar{z}}z-\dot{z}\bar{z}}{1+z\bar{z}}-i H(z,\bar{z}) \right)dt,
\end{array}
\end{equation}
where $\hbar = 1$ is assumed, $\xi_I$ and $\bar{\xi_F}$ are determined by the boundary conditions, and $H = \langle \mathcal{H}\rangle={\langle z|\mathcal{H}|z\rangle}/{\langle z|z \rangle}$ is the expectation value of the system's Hamiltonian $\mathcal{H}$. The quantity 
\begin{equation}
\mathcal{L}(z,\bar{z},\dot{z},\dot{\bar{z}})=S\frac{\dot{\bar{z}}z-\dot{z}\bar{z}}{1+z\bar{z}}-iH(z,\bar{z})
\end{equation}
must be viewed as the Lagrangian of the system. It is easy to see that 
\begin{equation}
\pdv{\mathcal{L}}{z} =  S\frac{\dot{\bar{z}}+\bar{z}^2\dot{z}}{(1+z\bar{z})^2}-i\pdv{H}{z}, \quad
\dv{t}\pdv{\mathcal{L}}{\dot{z}}  =  S\frac{-\dot{\bar{z}}+\bar{z}^2\dot{z}}{(1+\dot{z}z)^2}.
\end{equation}
Therefore the Euler-Lagrange equations of motion for the stereographic projections of the spin states on a sphere are \cite{graefe}
\begin{equation}
\dot{z}  =  -i\frac{(1+z\bar{z})^2}{2S}\pdv{H}{\bar{z}}, \quad \dot{\bar{z}} =  i\frac{(1+z\bar{z})^2}{2S}\pdv{H}{z}. 
\end{equation}

In order to compare these equations with traditional classical equations of motion for the spin, a slightly different formalism should be introduced. For a Hamiltonian $\mathcal{H}=H'-iH''$ it requires that operators $H'$ and $H''$ be Hermitian ($H'=H'^\dagger$, $H''=H''^\dagger$) and $\mathcal{P}$ symmetric. For the time independent Hamiltonian with a discrete spectrum, the time evolution of the state can be expressed as 
\begin{equation}
|\psi\rangle(t)=\sum_n c_ne^{-H''_n t} e^{iH'_n t}|\varphi_n\rangle,
\end{equation}
where $\varphi_n$ are the eigenstates of $\mathcal{H}$, with $H'_n$ and $H''_n$ being real and imaginary parts of the eigenvalues. The generalized Heisenberg equation of motion for the expectation value, $\langle \hat{O}\rangle={\langle z|\hat{O}|z\rangle}/{\langle z|z\rangle }$, of an arbitrary operator $\hat{O}$ is \cite{graefe}
\begin{equation}\label{evo}
i\dv{t}\langle \hat{O}\rangle=\langle [\hat{O},H]\rangle -2i\Delta_{\hat{O},H''}^2,
\end{equation}
where $\Delta_{\hat{O},H''}^2=\langle\{\frac{1}{2}\hat{O},H''\}\rangle-\langle\hat{O}\rangle\langle H''\rangle$, in which $\{,\}$ stands for the anti-commutator. For, e.g., $H'=aS_i+bS_j^2$ the first term in the right-hand-side of Eq.\ (\ref{evo}) is
\begin{equation}\label{commutator}
\langle [S_i,aS_j+bS_k^2]\rangle=i\varepsilon_{ijk}a\langle S_k\rangle+i\varepsilon_{ilm}b\langle\{S_l,S_m\}\rangle,
\end{equation} 
with $\langle \{S_l,S_m\}\rangle$ reducing to $2(1-\frac{1}{2S})\langle S_l\rangle\langle S_m\rangle+\delta_{lm}S$ in the limit of large $S$. In that limit (\ref{commutator}) coincides with \linebreak $(\nabla_S H)\times \bm{S}$.

For the second term in the right-hand-side of Eq.\ (\ref{evo}), when the non-Hermitian part of the Hamiltonian is chosen linear on the spin, $H'' = cS_k$, one obtains
\begin{equation}
\begin{array}{ll}
\Delta_{\hat{S_i},cS_k}^2&=\left\langle\left\{\frac{1}{2}\hat{S_i},cS_k\right\}\right\rangle-\langle\hat{S_i}\rangle\langle cS_k\rangle =\\ & = -\frac{c}{2S}\langle S_i\rangle\langle S_k\rangle+\delta_{ik}cS
\end{array}
\end{equation}
which coincides with $\frac{1}{S}[(\nabla_S H'')\times\bm{S}]\times{\bm{S}}$. Finally, one obtains
\begin{equation}\label{dyn}
\langle\bm{\dot{S}}\rangle=(\nabla_S \langle H' \rangle)\times \langle\bm{S}\rangle+\frac{1}{S}[(\nabla_S\langle H'' \rangle)\times\langle\bm{S}\rangle]\times{\langle \bm{S}\rangle}
\end{equation}

Consider now a non-Hermitian $\mathcal{PT}$-symmetric spin Hamiltonian 
\begin{equation}\label{PT}
\mathcal{H}_\mathcal{PT}= k_zS_z^2+i\beta S_y, 
\end{equation}
in which $k_z < 0$ describes an uniaxial crystal field (magnetic anisotropy) and $\beta$ is a real parameter.  This corresponds to the choice $H'=k_zS_z^2$ and $H'' =-\beta S_y$. In this case Eq.\ (\ref{dyn}) gives the following equations of motion for {\bf s} = {\bf S}/S:
\begin{equation}\label{mot}
\hbar \dot{\bf s} = 2k_z (\hat{z} \times {\bf s}) + \beta (\hat{y} \times {\bf s})\times {\bf s},
\end{equation}
where $\hbar$ has been restored. 

The LLS equation in the case of the uniaxial anisotropy and the electric current $I$ having spin polarization along the $y$-axis, that interacts via exchange with the localized spin ${\bf S}$, is given by \cite{Slonczewski96,Ralph,Tse,Cai}
\begin{equation}\label{LLS}
\hbar \dot{\bf s} = 2k_z \hat{z} \times {\bf s} +\frac{\hbar  \eta I} {2eS} (\hat{y} \times {\bf s})\times {\bf s} + \alpha (2k_z \hat{z} \times {\bf s}) \times {\bf s},
\end{equation}
where $\eta$ represents the degree of the spin polarization of the current, $0 < \eta <1$, and  $\alpha \ll 1$ is a dimensionless damping parameter. While the last term in Eq.\ (\ref{LLS}), that describes dissipation, can also be obtained with the use of a non-Hermitian Hamiltonian \cite{Wieser} such a Hamiltonian would not be $\mathcal{PT}$-symmetric and would not possess real eigenvalues regardless of the value of $\alpha$. In what follows we shall assume that the characteristic rate of the evolution of quantum spin states is much greater than the damping rate $\sim \alpha |k_z|/\hbar$ and will neglect contribution of the damping to the quantum dynamics. Comparison of Eq.\ (\ref{mot}) with Eq.\ (\ref{LLS}) immediately gives the relation suggested by Galda and Vinokur \cite{galda}: 
\begin{equation}\label{beta-I}
\beta = \frac{\hbar \eta}{2eS} I, 
\end{equation}
It forms the basis for the evaluation of the non-equilibrium effect of the spin-polarized current on a localized spin within $\mathcal{PT}$-symmetric quantum mechanics. While the relation (\ref{beta-I}) has been derived assuming $S \gg 1$ (which is true for many SMM) it must apply, up to a factor of order unity, to any spin.

%%%%%%%%%%%%%%%%%%%%%%%%%%%%%%%%%%%%%%%%%%%%%%%%%%%%%%%%%%%%%%%%%%%%%%%%%%%
\section{Non-Equilibrium Quantum Dynamics of a Localized Spin}

%%%%%%%%%%%%%%%%%%%%%%%%%%%%%%%%%%%%%%%%%%%%%%%%%%%%%%%%%%%%%%%%%%%
\subsection{Spin Tunneling in the Presence of Spin-Polarized Current} \label{tunneling}

For our purpose it is convenient to express the energy in the units of the anisotropy constant $|k_z|$ and to consider a non-Hermitian $\mathcal{PT}$-symmetric Hamiltonian
\begin{equation}\label{H-reduced}
\bar{\mathcal{H}}_\mathcal{PT}= -S_z^2 + \frac{1}{2}\bar{\beta}(S_+ - S_-)
\end{equation}
with $S_{\pm} = S_x \pm iS_y$, $\bar{\mathcal{H}}_\mathcal{PT} = \mathcal{H}_\mathcal{PT}/|k_z|$ and $\bar{\beta} = \beta/|k_z|$, that is equivalent to (\ref{PT}). The first term in Eq.\ (\ref{H-reduced}) creates a degeneracy for the eigenstates corresponding to the magnetic quantum numbers $\pm m$, while the second term removes that degeneracy. At $\bar{\beta} \ll S$ the eigenstates of the system can be formally studied by the perturbation theory for $\bar{\mathcal{H}}_\mathcal{PT} = H_0 + V$ with $H_0 = -S_z^2$ and $ V =  \frac{1}{2}\bar{\beta}(S_+ - S_-)$. This, of course, can only be fully justified if the resulting eigenstates are real. 

Small $\beta$ only weakly renormalize spin eigenstates, leaving their energies real. This is indicative of a static situation in which the spin-polarized current is too weak to generate any instabilities in the state of the system. For, e.g., $S = 1$, the splitting, $\Delta_{-1} = \bar{\beta}^2$, of the eigenvalue $E = -1$, caused by the perturbation, is determined by the secular equation
\begin{equation}
\left|\begin{array}{cc}
\frac{\bar{\beta}}{2}-E^{(2)}, & -\frac{\bar{\beta}^2}{2}\\
-\frac{\bar{\beta}^2}{2}, & \frac{\bar{\beta}}{2}-E^{(2)}
\end{array}\right|=0 \Longrightarrow E^{(2)}_+=0 \hspace{1mm};\hspace{1mm} E^{(2)}_-= {\bar{\beta}^2}
\end{equation}
and is real. Note that the matrix elements $V_{1,1}$ and $V_{-1,-1}$ are only responsible for the shift of the energy and do not contribute to $\Delta_{-1}$. 

\begin{figure}[h]
	\centering
	\includegraphics[width=\columnwidth]{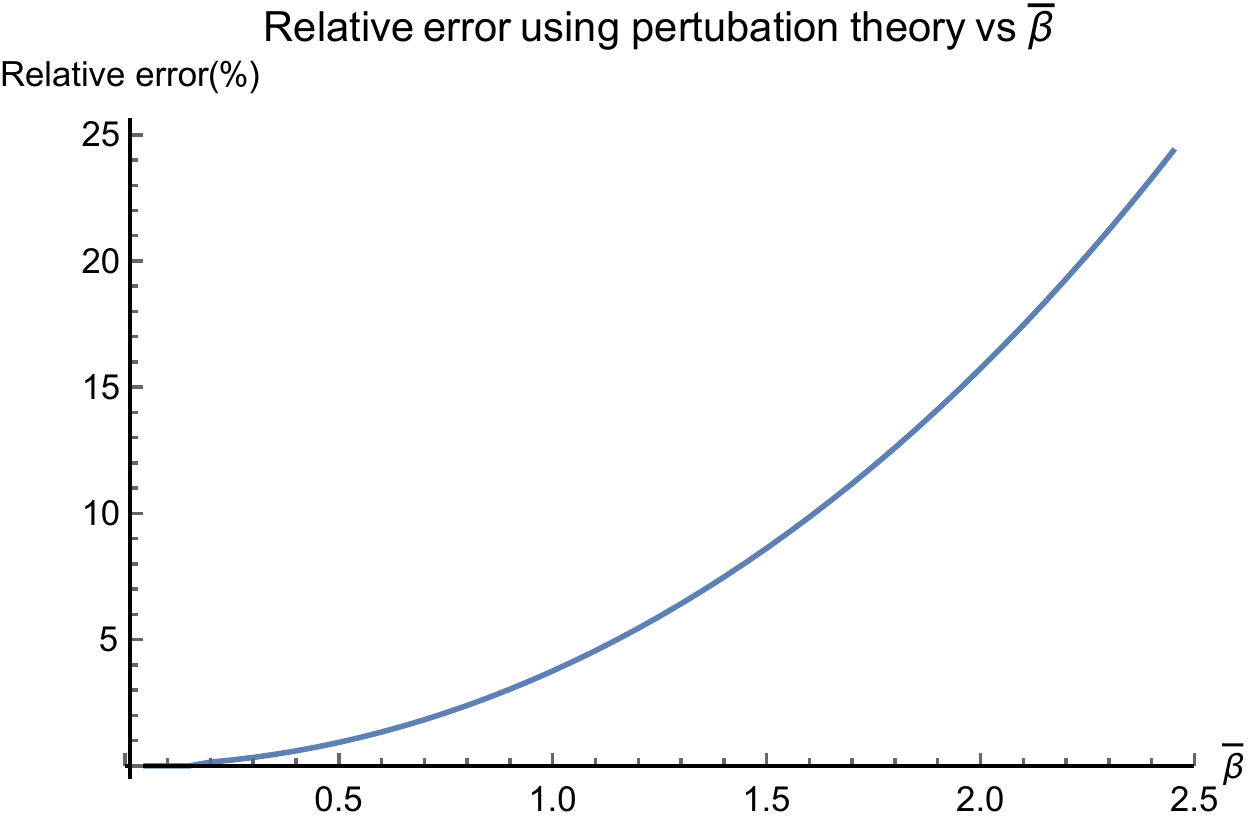}
	\caption{Percentage error of the perturbative result (\ref{pert}) for the ground state tunnel splitting $\Delta_{-5}$ as compared to the exact numerical diagonalization of the Hamiltonian (\ref{H-reduced}) for $S = 5$.}
	\label{fig: err}
\end{figure}
This appears to be a general situation for arbitrary $S$: The matrix elements $V_{i,i}$ do not contribute to the formula for the splitting of the $m$-th state $\Delta_m$,
\begin{equation}
\Delta_m=2\frac{V_{m,m+1}}{E_m^{(0)}-E_{m+1}^{(0)}}\frac{V_{m+1,m+2}}{E_m^{(0)}-E_{m+2}^{(0)}}\cdot \dots   V_{-m-1,-m},
\end{equation}
where $m<0$ has been assumed. Consequently, this formula provides the same real tunnel splitting as computed \cite{Garanin,Lectures} for a Hermitian Hamiltonian $\mathcal{H}_\mathcal{PT}= - S_z^2+\bar{\beta} S_y$:
\begin{equation}\label{pert}
\Delta_m = \frac{2}{[(-2m-1)!]^2}\frac{(S-m)!}{(S+m)!}\left(\frac{\bar{\beta}}{2}\right)^{2|m|}
\end{equation}

An independent check of the above perturbation result can be obtained by diagonalizing Hamiltonian (\ref{H-reduced}) numerically. The relative error of the formula (\ref{pert}), as compared to the exact numerical result, is shown for $S = 5$ in Fig.\ \ref{fig: err}.
As expected the deviation of the perturbation theory from the exact result is small for $\bar{\beta} \ll S$.

%%%%%%%%%%%%%%%%%%%%%%%%%%%%%%%%%%%%%%%%%%%%%%%%%%%%%%%%%%%%%%%%%%%%%%%%%%%
\subsection{Temporal Evolution of Spin States} \label{temporal}

\begin{figure}[ht]
	\centering
	\includegraphics[width=\columnwidth]{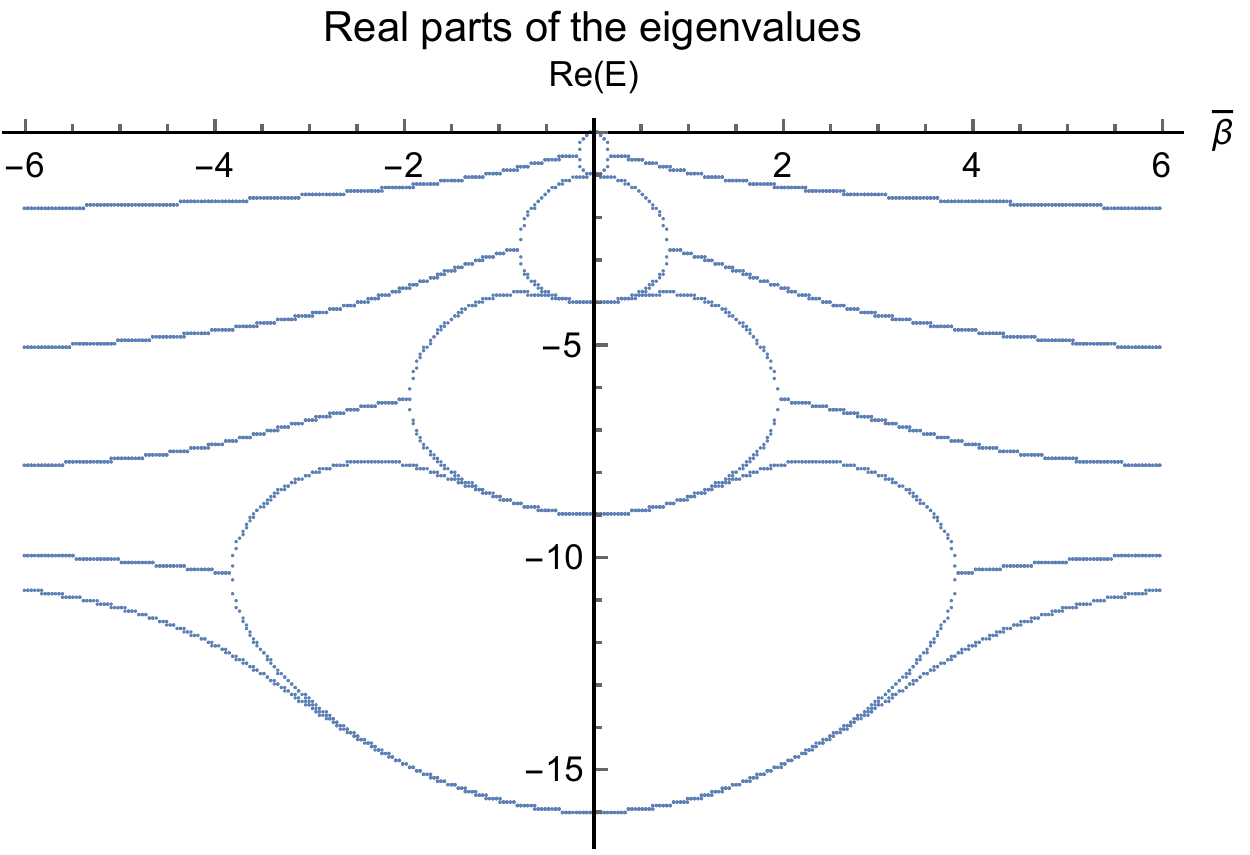}
	\includegraphics[width=\columnwidth]{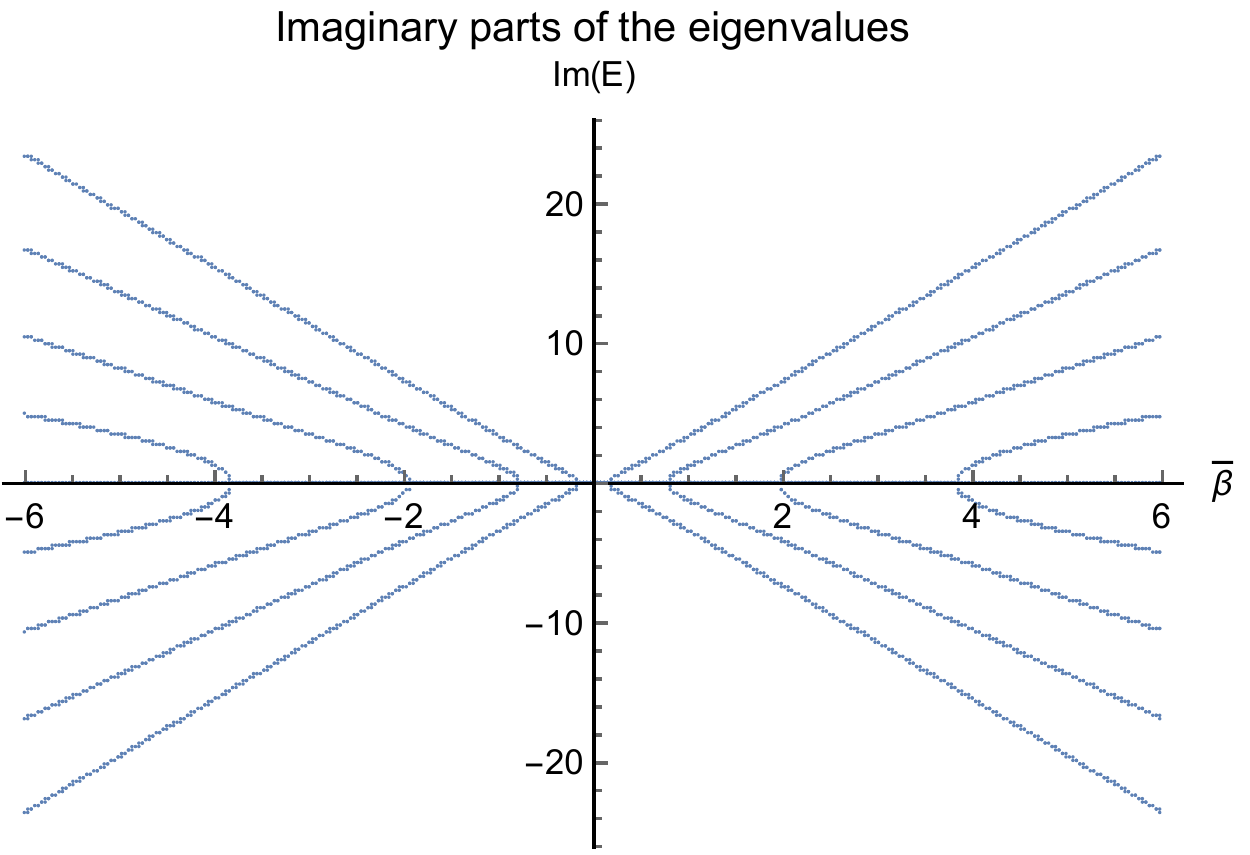}
	\caption{Upper panel: Real parts of the eigenvalues of the Hamiltonian (\ref{H-reduced}) for $S = 4$ as function of $\bar{\beta}$. Lower panel: The imaginary parts of the eigenvalues. New pairs of complex eigenvalues emerge at $\bar{\beta} = \pm 0.1636, \pm 0.7869, \pm 1.986. \pm 3.822$.}
	\label{fig:ERI}
\end{figure}

\begin{figure}[h]
	\centering
	\includegraphics[width=\columnwidth]{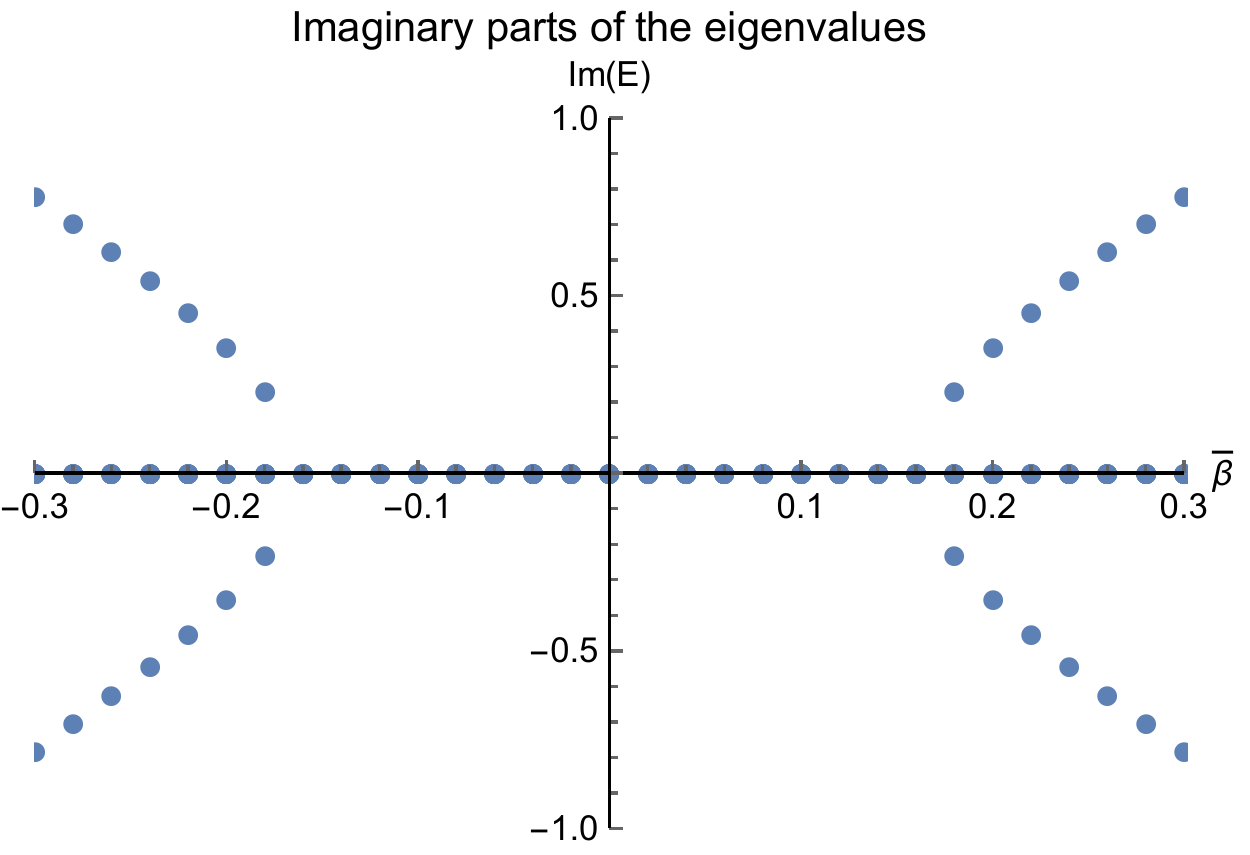}
	\caption{Amplification of the instability region in Fig.\ (\ref{fig:ERI}) near $\bar{\beta} = \bar{\beta}_{1}$ where the first pair of complex conjugate eigenstates emerges.}
	\label{fig:amplification}
\end{figure}

As $\bar{\beta}$ describing the effect of the spin-polarized current increases, pairs of complex conjugate energies begin to appear, starting with the smallest $|m|$. Further increase of $\bar{\beta}$ generates more complex conjugate pairs. This effectively corresponds to the tunnel splittings being consecutively switched from real to imaginary. The process ends when $S$ complex conjugate pairs emerge. This picture is illustrated in Fig.\ \ref{fig:ERI}. For an integer $S$ there is always one real energy because the total number of states is $2S + 1$. 
We will call the critical value of $\bar{\beta}$ at which the $n$-th pair of complex eigenstates emerges $\bar{\beta} = \bar{\beta}_{n}$. It corresponds to the critical current via the relation $I_{cn} = \frac{2eS|k_z|}{\hbar \eta}\bar{\beta}_{n}$. The area in Fig.\ (\ref{fig:ERI}) near $\bar{\beta} = \bar{\beta}_{1}$, where the first complex pair emerges, is amplified in Fig.\ \ref{fig:amplification}.

Complex energies $\varepsilon_n$ have profound consequences for the temporal evolution of the eigenstates given (in units of $\hbar=1$) by
\begin{equation}
|\varphi_n\rangle(t)=e^{-i\varepsilon_n t}|\varphi_n\rangle,
\end{equation}
One such consequence is the loss of normalization since for complex $\varepsilon_n$ the condition $|e^{i\varepsilon_nt}|=1$ is no longer satisfied. Another consequence is that occupation of the states with $\Im(\varepsilon_n) > 0$ will grow with time while occupation of the states with $\Im(\varepsilon_n) < 0$ will decrease. This determines the time evolution of the spin state that we will be addressing below.  
Writing for the eigenvalues $\mathcal{E}_n=E_n+i\Gamma_n$ an arbitrary spin state can be presented as $|\psi(t)\rangle\propto \sum_n c_n e^{\Gamma_n t}e^{-iE_nt}|\varphi_n\rangle$, where $|\varphi_n\rangle$ are the eigenstates of the Hamiltonian $\mathcal{H}_\mathcal{PT}$ and $c_n$ are arbitrary. The normalized states are given by
\begin{equation}\label{evolucio}
|\psi(t)\rangle=\frac{\sum_n c_n e^{\Gamma_n t}e^{-iE_nt}|\varphi_n\rangle}{\sqrt{\sum_n |c_n|^2e^{2\Gamma_nt}}}  
\end{equation}
We shall use the basis $|S,m\rangle$. Starting with one of the basis states, it is interesting to study how it evolves with time in the presence of the spin-polarized current, that is, at non-zero $\beta$. 

\begin{figure}[h]
	\centering
	\includegraphics[width=\columnwidth]{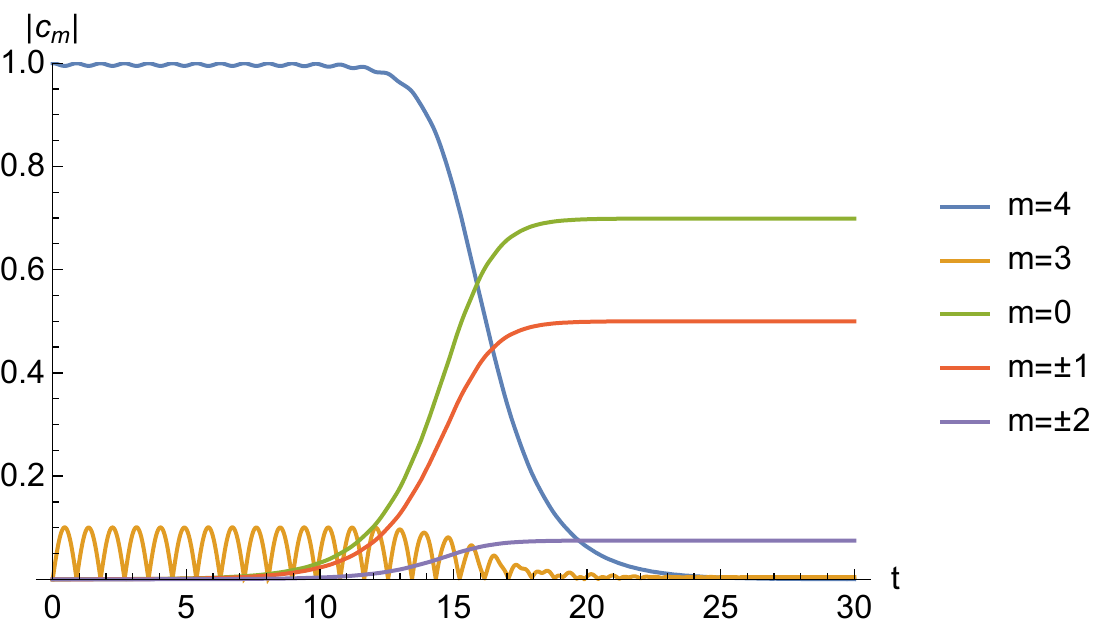}
	\caption{Time evolution of the spin system prepared in a state  $|\psi\rangle=|4,4\rangle$ for $\bar{\beta} = 0.25$. The lines show time dependence of the squere root of probabilities, $|c_m|$, in Eq.\ (\ref{cm}). Probabilities of the spin states that are not shown are close to zero.}
	\label{fig: dc}
\end{figure}

Evolution of the state 
\begin{equation}\label{cm}
|\psi(t)\rangle=\sum_mc_m(t)|4,m\rangle, \qquad \sum_m |c_m|^2 = 1
\end{equation}
with $\bar{\beta} = 0.25$ and the intial condition $|\psi(0)\rangle = |4,4\rangle$ is shown in Fig. \ref{fig: dc}. 
One interesting observation is a significant contribution of the states with negative $m$  to the final state of the system. In this paper we do not introduce interactions of the spin with other microscopic degrees of freedom that may cause dissipation. In the presence of the dissipation, once the spin transits from positive to negative $m$, it will travel down the energy staircase, thus completing the reversal from $|S,S\rangle$ to $|S,-S\rangle$. 

\begin{figure}[h]
	\centering
	\includegraphics[width=\columnwidth]{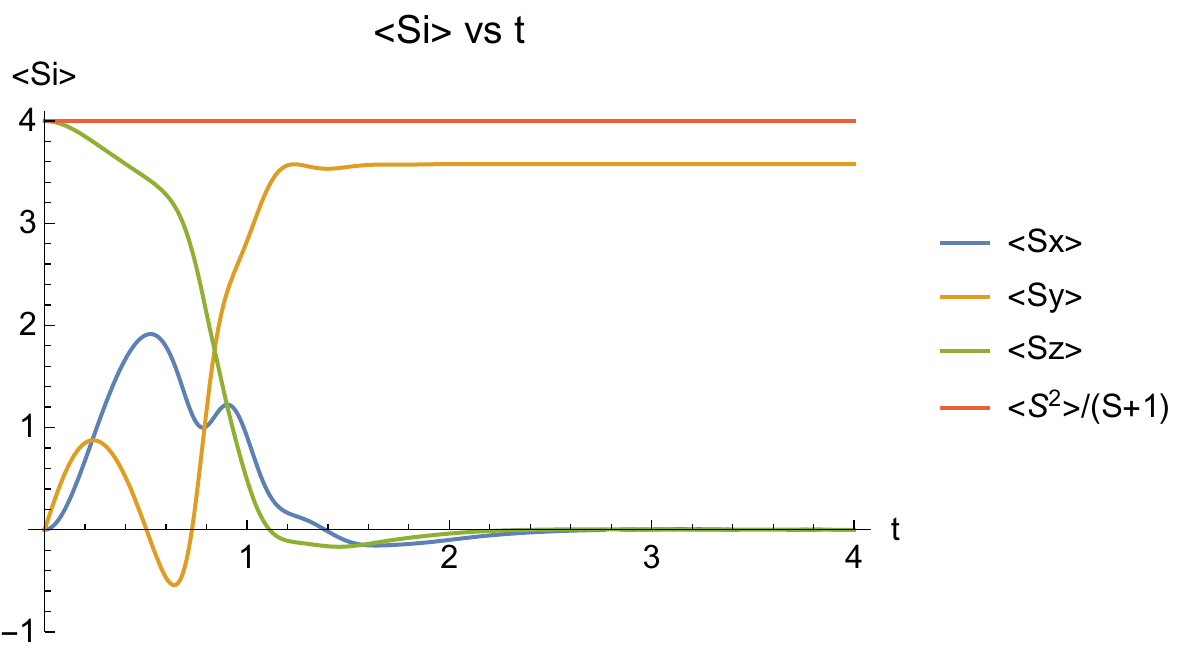}
	\caption{Evolution of the expectation values of the components of spin $S = 4$ from the intial state $|4,4\rangle$ caused by the spin-polarized current with $\bar{\beta} = 1.5$.}
	\label{fig:sz1}
\end{figure}

There is also another side to the story that reveals itself in the time evolution of the expectation values of the spin components $S_x$, $S_y$, and $S_z$, which is shown in Fig. \ref{fig:sz1}. As is seen in the figure the length of the spin is preserved due to the condition $\hat{\bf S}^2 = S(S+1)$. From a classical point of view, however, the effect of the spin-polarized current consists of the rotation of the localized spin from its initial orientation along the $z$-axis to the orientation along the $y$-axis. Notice also that the $S_z$ component briefly crosses to the negative territory, that is, the spin goes over the anisotropy energy barrier, which must be sufficient to achieve full reversal in the presence of dissipation. 

%%%%%%%%%%%%%%%%%%%%%%%%%%%%%%%%%%%%%%%%%%%%%%%%%%
\section{Spin-reversal rate} \label{reversal}

For small $\bar{\beta} < \bar{\beta}_c$ all $\Gamma$ are zero and the degeneracy of each pair of states $|S,m\rangle$ and $|S,-m\rangle$ is removed by the real splitting due to the quantum tunneling between these states. As is well known, prepared in a state $|S,m\rangle$ the spin will oscillate between $|S,m\rangle$ and $|S,-m\rangle$ at a frequency $\Delta_m/\hbar$. On the contrary, at $\bar{\beta} > \bar{\beta_c}$ the ``splitting'' of some pairs becomes imaginary with opposite signs of $\Gamma$ for the states that evolve from $|S,m\rangle$ and $|S,-m\rangle$ and have the same real part of the energy. These states correspond to spin up and spin down on two sides of the energy barrier determined by the magnetic anisotropy. In the case of a non-zero $\Gamma$ (imaginary ``splitting'') the occupation numbers of the states with negative $\Gamma$ on one side of the energy barrier will exponentially go down, while occupation numbers of the states with positive $\Gamma$ on the other side of the barrier will exponentially increase, providing the reversal of the spin. 
\begin{figure}[h]
	\centering
	\vspace{0.5cm}
	\includegraphics[width=\columnwidth]{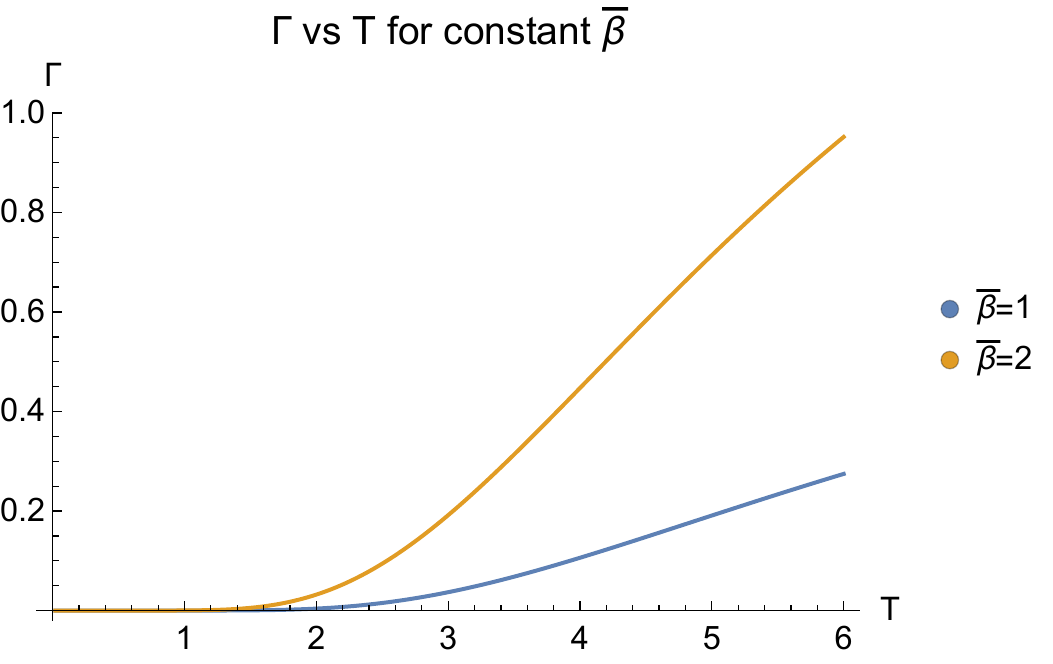}
	\includegraphics[width=\columnwidth]{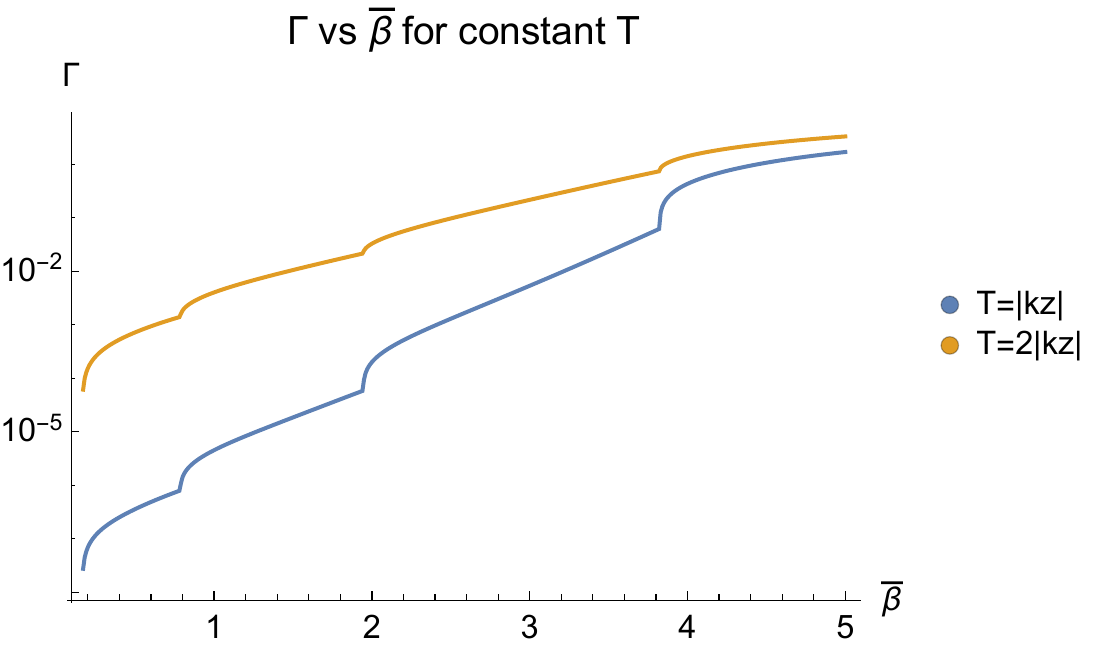}
	\caption{Upper panel: Temperature dependence of $\Gamma$ for the Hamiltonian (\ref{H-reduced}) at $S = 4$ and $\bar{\beta} = 1$ and $2$ (T is measured in the units of $|k_z|$.) Lower panel: The dependence of $\hat{\Gamma}$ on $\bar{\beta}$ at a constant temperature $T= |k_z|$ and $T= 2|k_z|$.}
	\label{fig:Gamma}
\end{figure}
With the above picture in mind one can define the rate of the spin reversal in a conventional manner:
\begin{equation}
{\Gamma}(\beta, T)=\frac{\sum_n |\Gamma_n| e^{-\frac{ E_n}{T}}}{\sum_n e^{-\frac{ E_n}{T}}}
\end{equation}
with $T$ being the absolute temperature. Here we measure all quantities (${\Gamma}, \Gamma_n, E_n, T$) in the units of the anisotropy constant $|k_z|$. The latter is typically well below 1K.  The dependence of ${\Gamma}$ on $T$ and $\bar{\beta}$ for the Hamiltonian (\ref{H-reduced}) at $S = 4$ is shown in Fig.\ \ref{fig:Gamma}. While the $T$-dependence of ${\Gamma}$ is smooth, the dependence on $\bar{\beta}$ shows kinks associated with the emergence of pairs of complex eigenvalues. The first two pairs emerge at $\bar{\beta}_1 \approx 0.16$ and $\bar{\beta}_2 \approx 0.79$. Any $\bar{\beta} > \bar{\beta}_1$ causes instability manifested by a finite ${\Gamma}$. At $T =|k_z|$ and $\bar{\beta} < 1$ however the rate is exponentially small to be distinguished from zero in Fig.\ \ref{fig:Gamma}. The next two pairs of complex eigenvalues emerge at $\bar{\beta}_3 \approx 2.0$ and $\bar{\beta}_4 \approx 3.8$. They are clearly seen as the kinks in Fig.\ \ref{fig:Gamma}  that result in the rise of $\Gamma$.

%%%%%%%%%%%%%%%%%%%%%%%%%%%%%%%%%%%%%%%%%%%%%%%%%%%%%%%%%%%%%%%%%%%%%%%%%%%
\section{Discussion} \label{discussion}
We have proposed a quantum processor in which magnetic qubits are manipulated by a spin-transfer torque, see Fig.\ \ref{fig: device}. The effect of the spin-polarized current on the quantum states of a localized tunneling spin of, e.g., a single-molecule magnet (SMM) has been described within the approach based upon $\mathcal{PT}$-symmetric quantum mechanics. A sufficiently weak current only weakly perturbs the quantum spin states. In a manner similar to the magnetic field, it provides the splitting, $\Delta_m$, of the $|\pm m\rangle$ states that were degenerate in the absence of the current. At some critical value of the current, $I = I_{c1}$, the real part of the splitting of the states at the top of the anisotropy barrier becomes zero, but the corresponding degenerate eigenvalues acquire imaginary parts of opposite sign. On increasing the current the same happens to the tunnel splittings of the lower energy states all the way down to the ground state. 

The physical picture associated with the above mathematics of the $\mathcal{PT}$-symmetric spin Hamiltonian naturally corresponds to the instability caused by the spin-polarized current: Population of the spin states on one side of the anisotropy energy barrier begins to grow, while population of the states on the other side of the barrier begins to collapse, leading to the spin reversal induced by the current. The corresponding rate $\Gamma$ depends on temperature and the magnitude of the current in a non-trivial way. At low temperature and the current just above $I_{c1}$ the spin states at the top of the barrier are not occupied and the effect of the current is exponentially weak. It grows with temperature exponentially in a continuous manner. It also increases with the magnitude of the current via kinks seen in the dependence of $\Gamma$ on $I$ at $I = I_{cn}$. These values of the current are related to the critical values of the dimensionless parameter $\bar{\beta}$, via  
\begin{equation}\label{I-n}
I_{cn} = \frac{2eS|k_z|}{\hbar \eta}\bar{\beta}_n,
\end{equation}
where $|k_z|$ is the magnetic anisotropy constant and $\eta$ is the degree of the polarization of the current. For e.g., $S = 4$ the first critical values of $\bar{\beta}$ is  $\bar{\beta}_1 = 0.1636$. 

Choosing for example $S = 4$, $|k_z| = 0.1$K, $\eta = 1$ we obtain from Eq.\ (\ref{I-n}) $I_{c1} \approx 0.7 $nA. This value of the current is typical in experiments with the electronic transport through a molecule bridged between two conductors, such as, e.g., an STM tip and a substrate in a single-molecule transistor setup \cite{Kai,Perrin}. We therefore conclude that manipulation of the spin states of an SMM by a spin-polarized electric current is within experimental reach. It can serve as a working concept for a quantum processor, with algorithms described by $\mathcal{PT}$-symmetric quantum mechanics. Quantum gates based upon such principle would be scalable in, e.g., a device schematically shown in Fig.\ \ref{fig: device}. Its advantage is selective manipulation of individual nanoscale qubits as compared to the spread-out effect of the external magnetic field. The speed of the quantum processor controlled by the spin-polarized electric current would also be much higher than the speed of the device controlled by the magnetic fields.

\section{Acknowledgments}
The authors are grateful to Alexey Galda and Valerii Vinokur for introducing them to the problem of spin dynamics within $\mathcal{PT}$-symmetric quantum mechanics. This work has been supported by the Grant No. DEFG02-93ER45487 funded by the U.S. Department of Energy, Office of Science.

\end{document}